# Superconducting properties in tantalum decorated three-dimensional graphene and carbon structures

Cayetano S.F. Cobaleda[1,2,*], Xiaoyin Xiao[1], D. Bruce Burckel[1], Ronen Polsky[1], Duanni Huang[1,3], Enrique Diez[2] and W. Pan[1,*]

[1]Sandia National Laboratories, P.O. Box 5800, MS 1086, Albuquerque, New Mexico 87185, U.S.A

[2]Laboratorio de Bajas Temperaturas, Universidad de Salamanca, E-37008 Salamanca, Spain

[3]Department of Electrical and Computer Engineering, University of California, Santa Barbara, California 93106, U.S.A.

ABSTRACT

We present here the results on superconducting properties in tantalum thin films (100nm thick) deposited on three-dimensional graphene (3DG) and carbon structures. A superconducting transition is observed in both composite thin films with a superconducting transition temperature of 1.2K and 1.0K, respectively. We have further measured the magnetoresistance at various temperatures and differential resistance dV/dI at different magnetic fields in these two composite thin films. In both samples, a much large critical magnetic field (~ 2 Tesla) is observed and this critical magnetic field shows linear temperature dependence. Finally, an anomalously large cooling effect was observed in the differential resistance measurements in our 3DG-tantalum device when the sample turns superconducting. Our results may have important implications in flexible superconducting electronic device applications.

Research on two-dimensional (2D) graphene has attracted enormous interests since the integer quantum Hall effect was observed in this material system in 2005 [1,2]. Due to the nature of an exposed 2D electron system open to environment, many studies have been carried out in examining how adsorption of metal adatoms on graphene sheet can modify its electronic transport properties for electronics, sensors and energy applications [3-6]. Indeed, in tin decorated graphene sheets a tunable superconducting phase transition was observed [7]. There, the non-percolating tin clusters dope the graphene sheet and induce long-range superconducting correlations. Moreover, it was shown that lithium decorated graphene can display a much higher superconducting transition temperature than in the superconductivity in bulk graphite intercalated by lithium atoms [8].

In recent years, three-dimensional graphene structures [9-12], a flexible and conductive interconnected graphene network, have generated a great deal of interests. Compared to 2D graphene sheets, surface area in 3D graphene structures is greatly enhanced. Furthermore, 3D graphene maintains the outstanding electrical, thermal, and mechanical properties as in 2D graphene. As a result, 3D graphene structures are expected to play an important role in flexible electronics, sensors, and energy storage applications (e.g., supercapacitors, battery electrodes, and hydrogen storage).

In this Letter, we present our recent electronic transport results on superconducting properties in tantalum decorated 3D graphene and carbon structures, following the pioneering work of decorating 2D graphene sheets with superconducting materials. A superconducting transition is observed in both composite thin films with a superconducting transition temperature of ~1.2 and 1K, respectively. The magnetoresistance at various temperatures and differential resistance dV/dI at different magnetic fields were also carried out. The obtained critical magnetic field shows linear temperature dependence. Moreover, an anomalously large cooling effect was observed in the differential resistance measurements in our 3D graphene - tantalum composite when the device turns superconducting.

The 3D carbon (3DC) structures were prepared from photoresist spun on silicon wafers using interference lithography, followed by pyrolysis and graphitic conversion via nickel catalyst [12].

Figure 1(a) shows the carbon structure consisting of five interconnected layers of ~800 nm hexagonal pores. The connecting arms are ~80 nm in diameter in the top layer and ~200 nm in the bottom layers. The carbon in this structure is mostly SP$^3$ and amorphous. A smooth, nominally conformal Ta coating on the carbon is shown in Figure 1(b). The conversion of SP$^3$ carbon into SP$^2$ carbon was done through the rapid thermal annealing in the presence of Ni catalyst [12]. Ni was removed by sulfuric acid. Figures 1(c) and 1(d) show their structures before and after Ta coating. Noticeably, a few layers of graphene with typical wrinkle characteristics were formed, and the arms and nods are largely expanded. The pattern is still hexagonal.

The 3DG-Ta and 3DC-Ta samples we examined both have a size of ~ 4mm×4mm. In each sample, eight ohmic contacts, made of silver epoxy, were placed symmetrically around the sample perimeter. An elevated temperature of 60 degrees was used for a fast epoxy curing. A schematic of the sample with contact arrangement is shown in the inset of Figure 2. The data were collected using the standard low frequency (~ 45 Hz) phase lock-in techniques.

In Figure 2, we show the four-terminal resistance $R_{xx}$ as a function of temperature in the 3DG-Ta sample. A superconducting transition is clearly seen. The resistance, ~ 60 ohms, is relatively flat at high temperatures. It increases a little bit before it drops abruptly to zero at low temperatures, a signature of superconducting transition. Following the standard definition, we determined the superconducting temperature, $T_c$ ~ 1.2K, as the point at the half of the normal resistance. We note that this transition temperature is significantly smaller than that in bulk Ta (~ 4.47K), due to the disordered nature in our Ta film. The width of the superconducting transition is $\Delta T_c$ ~ 150mK. It is significantly larger than that in bulk Ta (~ 4mK) [13], again due to a disordered Ta film. The superconducting transition temperature determined in this figure is the same as the one from critical magnetic field measurements (shown in Figure 3(b)). For comparison, a superconducting transition in a 3D carton coated with 100nm Ta thin film was also observed. The transition temperature, obtained from critical magnetic field measurement (as shown in Figure 3(d)), is a little bit lower, ~ 1K.

A higher $T_c$ in 3DG-Ta than in 3DC-Ta is unexpected. First, the Ta films were deposited on 3DG and 3DC at the same time under the same conditions. As a result, the variation in Ta film thickness can be excluded for the difference in $T_c$. Secondly, by examining the SEM images in Figures 1(b) and 1(d), it is clearly seen that the Ta film on 3DC is much smoother than that on 3DG. Naively, one can expect that the quality of Ta thin film should be better in 3DC, and, therefore, a high $T_c$ is expected. This is not we observed. We believe that this apparent discrepancy is most-likely due to the difference in the underneath substrates (i.e., $SP^2$ graphene versus $SP^3$ carbon) and the so-induced charge transfer between Ta and 3DG and between 3DC-Ta. Indeed, it has been shown that the substrate plays an important role in the strength of the electron-phonon interactions in metal-decorated structures [14]. Moreover, the charge transfer between metal-adatoms and the substrate is quite different in graphene and graphite structures [3,8]. It is then possible that the hybridization between the electrons in Ta and electrons in $SP^2$ carbon is stronger than that between Ta and $SP^3$ carbon [15]. As a consequence, the density of state near the Fermi level is larger in 3DG-Ta, which gives rise to stronger electron-phonon interactions and, consequently, higher $T_c$. In addition, we also note that the difference in $T_c$ is even bigger in thinner Ta films deposited on 3DG and 3DC structures. Indeed, in the samples with a 20nm thick Ta film, a $T_c$ of ~ 0.75K was observed in 3DG-Ta while no superconducting transition in 3DC-Ta.

In Figure 3(a), we show the magneto-resistance at a few selected temperatures of 0.36, 0.63, 0.92, 1.23, 1.49, and 6.01K. Here, the magneto-resistance is normalized to that at higher magnetic fields where Ta is in the normal state. At 0.36K for 3DG-Ta, the resistance is zero around zero magnetic field. It increases quickly when B > 1.4T, and the resistance is roughly constant at high magnetic field when the Ta film is in the normal state. This set of data thus demonstrates a magnetic field induced destruction of superconducting state. The critical magnetic field $B_c$ is ~ 1.38T, defined by the largest magnetic field at which $R_{xx}/(R_{xx})_N$ is still zero. With increasing temperature, $B_c$ decreases. At T ~ 1.23K, The $B_c$ is already zero. In Figure 2(b), we plotted the $B_c$ as a function of temperature. In general, $B_c$ follows a linear dependence on T. We fit this linear temperature dependence using the standard Ginzburg-Landau theory for an isotropic superconductor, $H_{c2} = \phi_0/2\pi\xi(0)^2 \times (1-T/T_c)$ [16,17]. Here $\phi_0 = h/2e$ is the flux quantum, $\xi(0)$ is the superconducting coherence length at T = 0. From the fitting, we can deduce that $T_c$ is ~ 1.22K, consistent with that in the $R_{xx}$ measurement in Figure 2. $\xi(0)$ is ~ 14nm, much smaller than that (~ 93 nm) in bulk Ta, due to the disordered nature of the Ta film. We also note here that the deduced zero temperature critical magnetic field, ~ 2.0T, in this composite is much higher than that (~ 800 Gauss) in bulk Ta [18]. This large critical magnetic field is due to flux pinning in the disordered and inhomogeneously distributed Ta film deposited on the 3D structures.

For comparison, we show the results in 3DC-Ta in Figure 3(c) and 3(d). In general, the data in 3DC-Ta are very similar to those in 3DG-Ta, expected for a lower $T_c = 0.96$K (as obtained in the fitting in Figure 3(d)) and a higher $B_c = 2.3$T at T = 0.

Knowing $\xi(0)$, we can deduce the Fermi velocity ($V_F$) in both composites based on the formula $\xi(0) = \hbar V_F/(1.76\pi k_B T_c)$. $V_F$ is ~ $1.2\times10^4$ m/s in 3DG-Ta, high than that in 3DC-Ta (~ $0.8\times10^4$ m/s). We notice that $V_F$ in both 3DG-Ta and 3DC-Ta is lower than that (~ $3\times10^5$ m/s) in bulk Ta (assuming $\xi(0) = 93$nm and $T_c = 4.47$K).

Figure 4 shows the differential resistance dV/dI as a function of the DC bias in 3DG-Ta at three temperatures. Again, dV/dI is normalized to that at higher bias where Ta is in the normal state. In this kind of measurements, a sweeping DC bias ($V_{DC}$) plus a small AC bias (10µV) was applied between contact 1 and contact 5 in series of a 100-ohm resistor. dV/dI was measured between contact 2 and 4.

At 4.2K, well above the transition temperature of $T_c$ ~ 1.2K, dV/dI is constant over the whole bias range. As the sample temperature is lowered down to 1.5K, close to $T_c$, a weak dV/dI dip has appeared at the zero bias (inset of Figure 4). As the sample temperature is further lowered to ~ 0.4K, this dip becomes the zero between ~ -0.5 and 0.5V. Taking into account the contact resistance in our sample, we estimated critical current $I_C$ to be ~ 1 mA.

Finally, we report an anomalously large cooling effect in our 3DG-Ta device observed in the dV/dI measurements when the sample turns superconducting. In Figure 5, we plot the reading of a thermometer located very close to the 3DG-Ta device. Starting from the high bias ends, the fridge temperature decreases with decreasing $V_{DC}$ due to reduced sample heating. At $V_{DC}$ ~ ±0.5V where the sample turns superconducting (as shown in Figure 4), the fridge temperature drops precipitately from ~ 0.7K to ~0.55K before it gradually cools down to a base temperature of ~ 0.3K. This large cooling effect is unexpected. On the other hand, it could be due to a large superconducting tunneling junction area between 3DG and Ta thin films. We note here that this large cooling effect plus the nature of flexible 3DG thin film can be utilized for solid-state cooling [19] applied to irregular objects.

In conclusion, we have studied superconducting properties of the tantalum thin film deposited on 3D graphene and 3D carbon structures. A superconducting transition was observed in both samples, and the transition in the tantalum decorated 3D graphene takes place at higher temperature than in the tantalum decorated 3D carbon, suggesting a stronger coupling between Ta and $SP^2$ graphene. The critical magnetic field from the magnetoresistance measurements was determined to ~ 2T, much higher than the Ta bulk value. From the dV/dI measurements, a critical current of ~ 1mA was obtained. Finally, an anomalously large cooling effect was observed in our 3DG-Ta device. Our results are expected to have important implications in flexible graphene superconducting electronic devices applications.

This work was jointly supported by the U.S. Department of Energy, Office of Science, Basic Energy Sciences, Materials Sciences and Engineering Division (C.C., D.H., and W.P.) and by LDRD at Sandia (X.Y.X., D.B.B., R.P., and W.P.), and also by the Spanish MINECO MAT2013-46308-C2-1-R (C.C. and E.D.), FPU-ap2009-2619 (C.C.), and JCYL SA226U13 (C.C. and E.D.). Sandia National Laboratories is a multi-program laboratory managed and operated by Sandia Corporation, a wholly owned subsidiary of Lockheed Martin Corporation, for the United States Department of Energy's National Nuclear Security Administration under contract DE-AC04-94AL85000.

AUTHOR INFORMATION

**Corresponding Authors**

* Email addresses: ccobaleda@usal.es, wpan@sandia.gov

Figures and Figure Captions

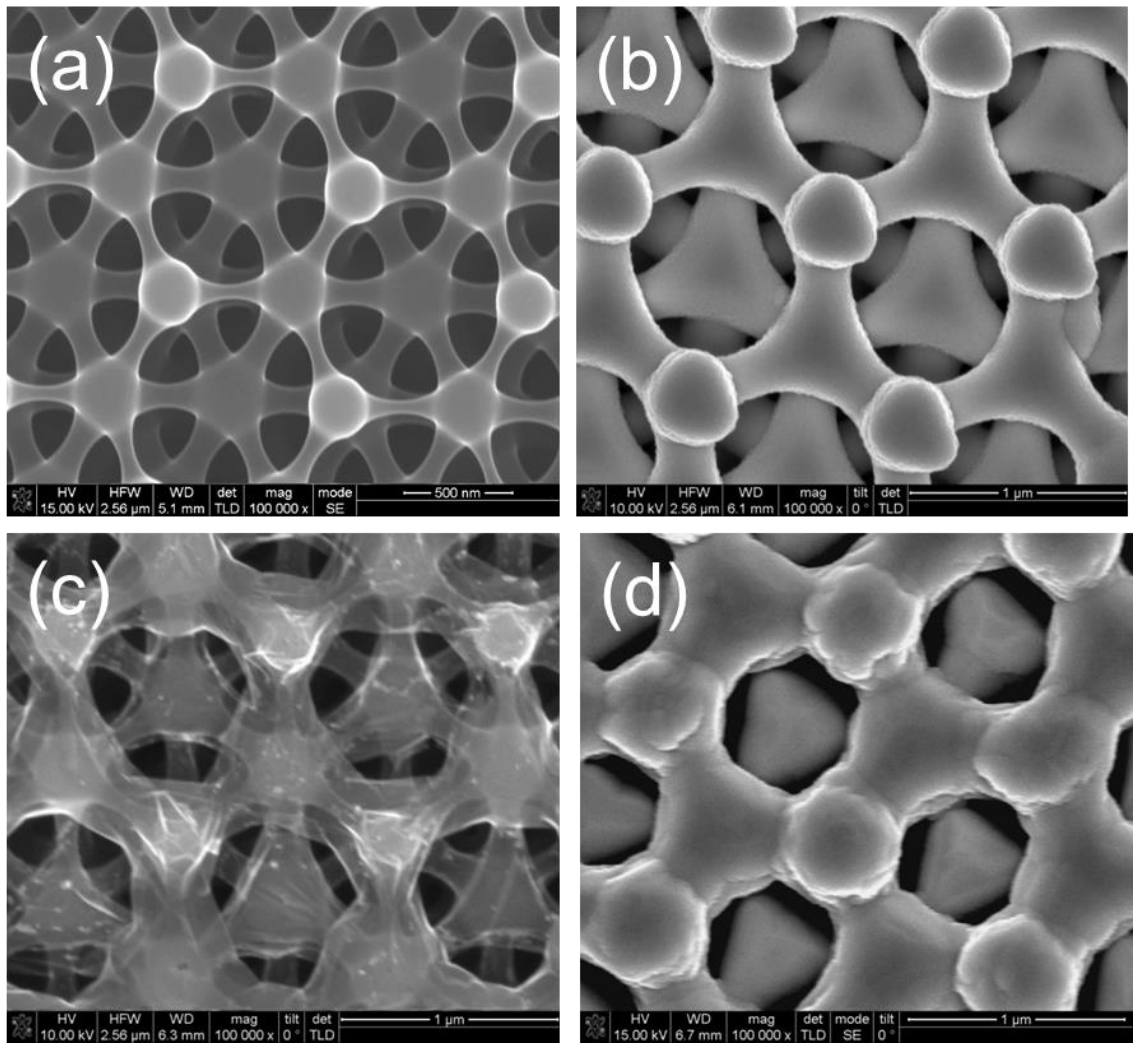

**Figure 1**. SEM images of 3D amorphous carbon (a) and 3D graphene (c) structures; as well as their corresponding Ta coated structures 3DC-Ta (b) and 3DG-Ta (d).

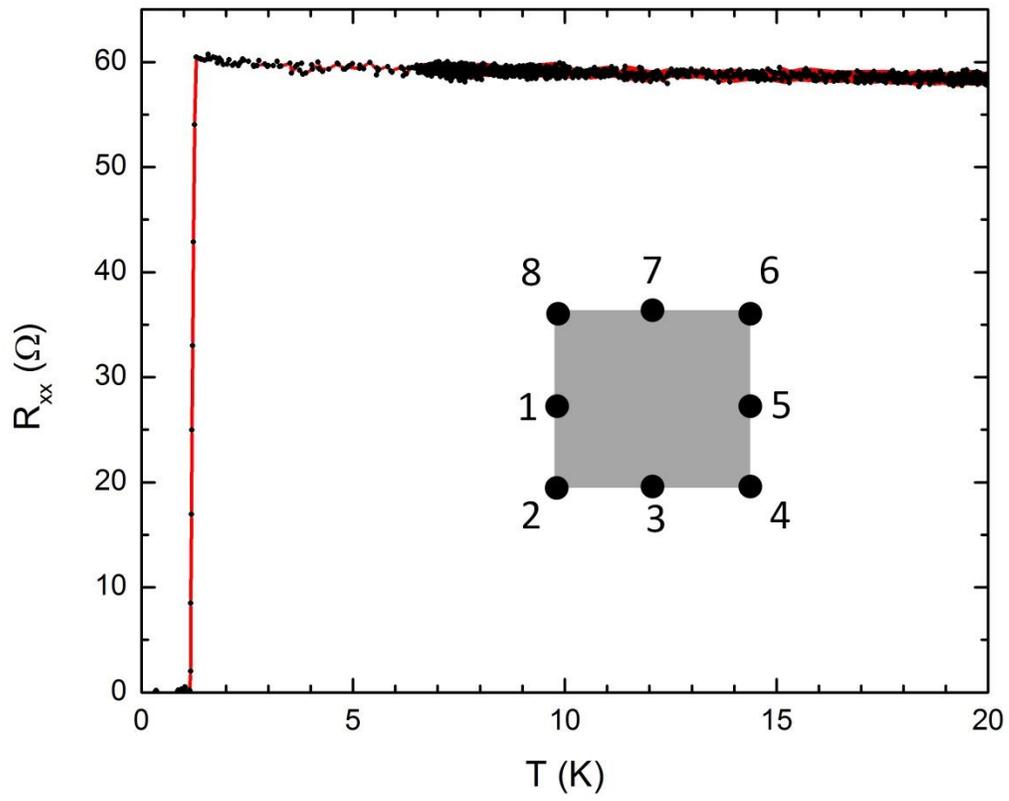

**Figure 2**. Four-terminal resistance $R_{xx}$ as a function of temperature at zero magnetic field. The inset shows a schematic of typical sample with contact arrangement.

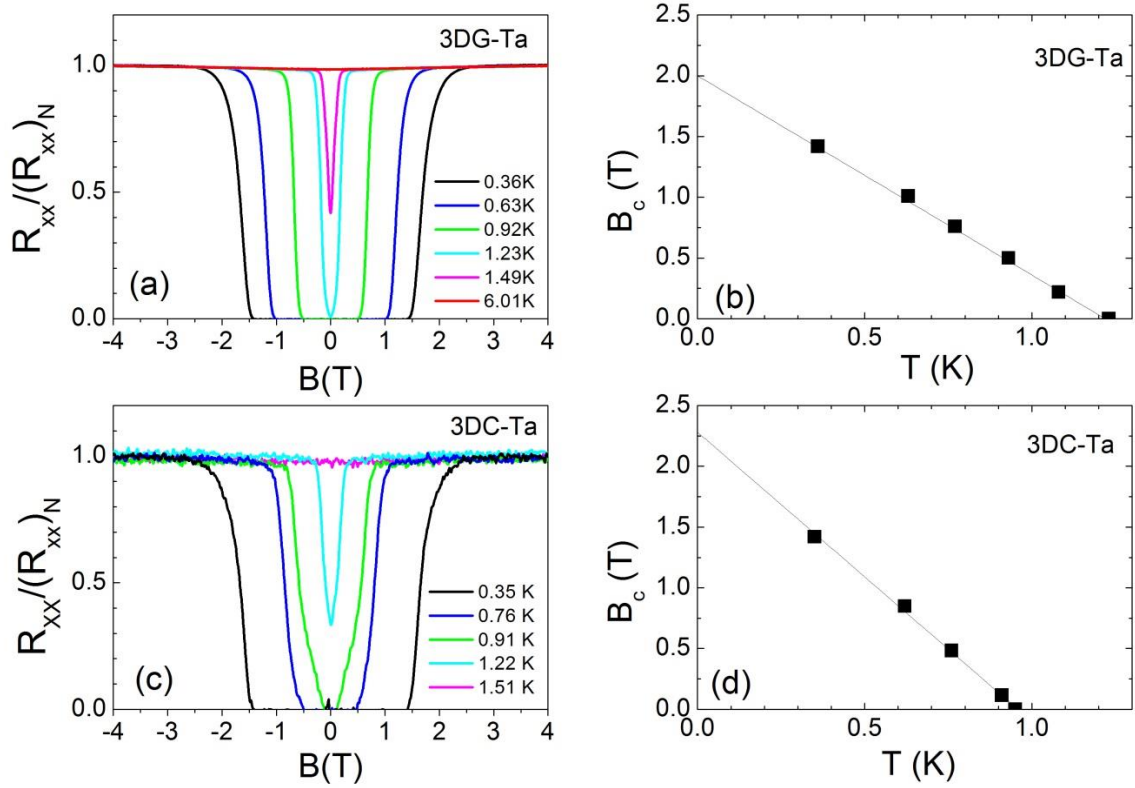

**Figure 3**. (a) and (c) show the normalized magneto-resistance at a few selected temperatures in 3DG-Ta and 3DC-Ta. (b) and (d) show $B_c$ as a function of temperature for 3DG-Ta and 3DC-Ta, respectively. Lines are linear fit.

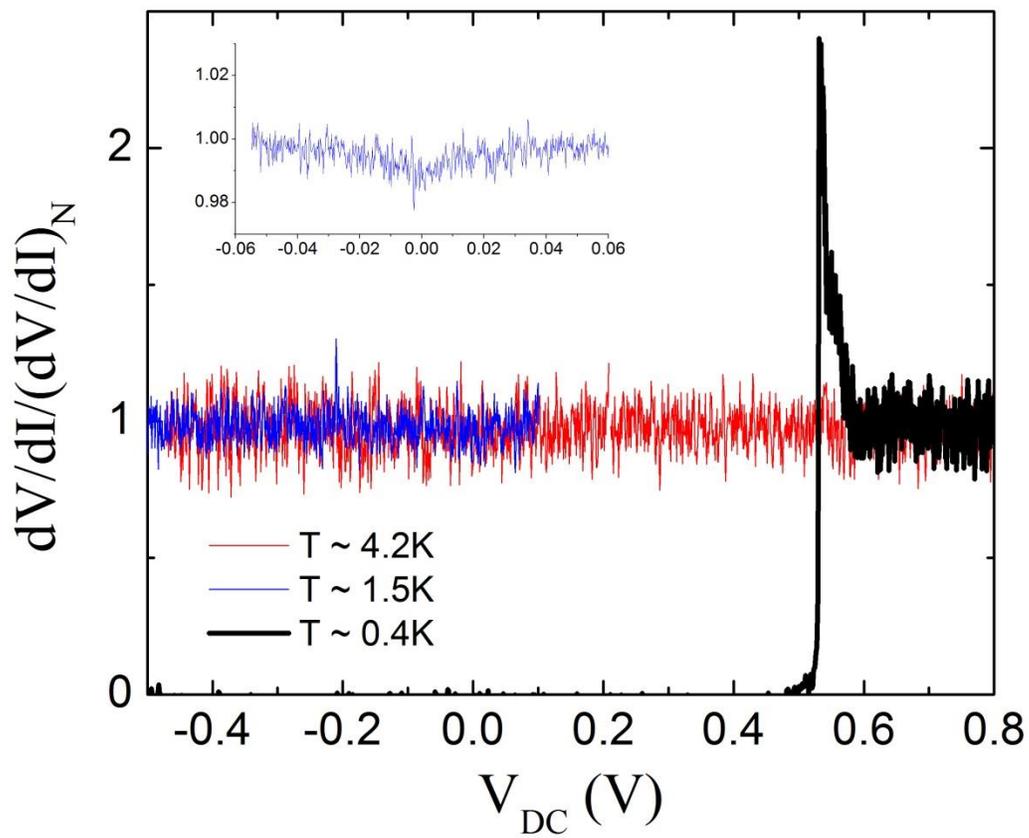

**Figure 4**. Normalized dV/dI in 3DG-Ta at T ~ 4.2, 1.5, and 0.4K.

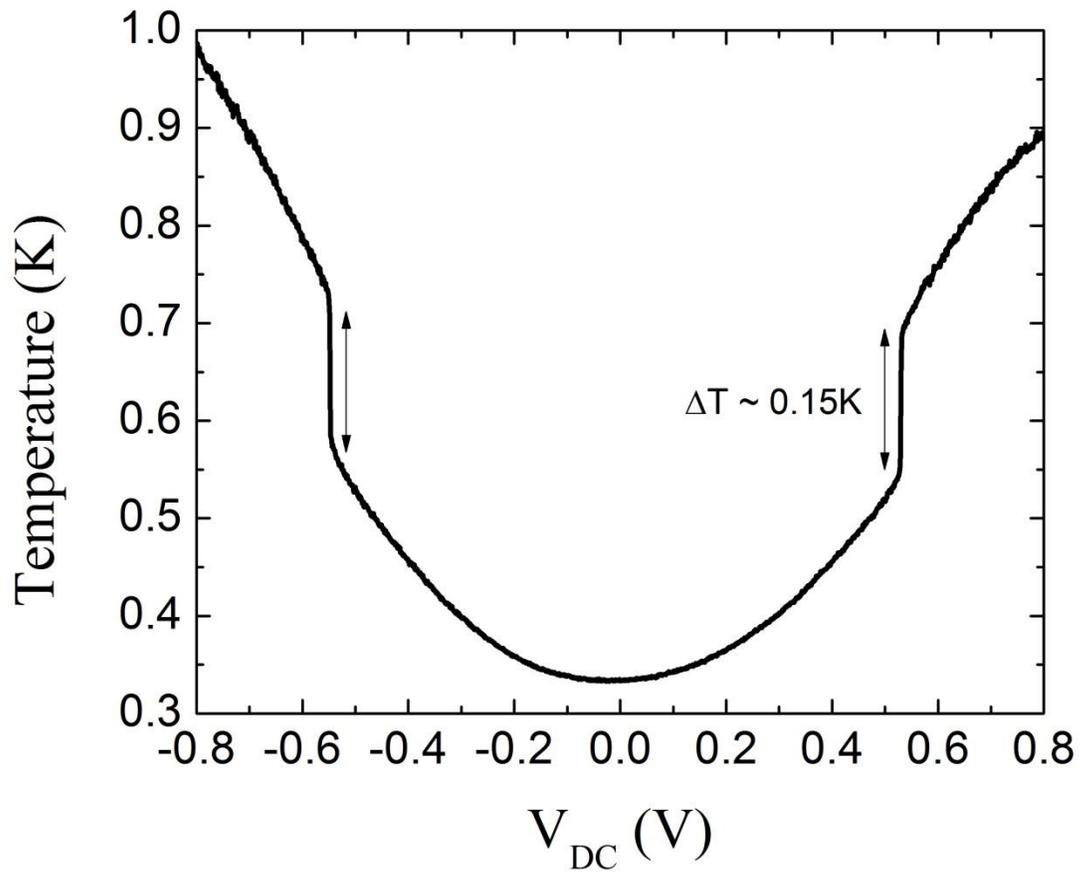

**Figure 5**. Large cooling effect observed in 3DG-Ta. The base temperature is ~ 0.3K.